\documentclass[amsmat,amssymb,amsfonts,aps,prb,twocolumn,showpacs]{revtex4}

\usepackage{graphicx}
\usepackage{dcolumn}
\usepackage{bm}
\begin{document}

\title{Orbital eigenchannel analysis for ab-initio quantum transport calculations}

\author{ David Jacob }
\email{david.jacob@ua.es}
\author{ J. J. Palacios}
\affiliation{Departamento de F\'isica Aplicada and Instituto Universitario de Materiales
de Alicante (IUMA), Universidad de Alicante, 03690 San Vicente del Raspeig, SPAIN }

\date{\today}


\begin{abstract}
We show how to extract the orbital contribution to the transport eigenchannels from 
a first-principles quantum transport calculation in a nanoscopic conductor.
This is achieved by calculating and diagonalizing the first-principles transmission matrix 
reduced to selected scattering cross-sections.
As an example, the orbital nature of the eigenchannels in the case of Ni nanocontacts 
is explored, stressing the difficulties inherent to the use of non-orthogonal basis sets 
\end{abstract}

\maketitle

\section{Introduction}
\textit{Ab initio} quantum transport calculations for nanoscopic conductors like 
molecular junctions and nanocontacts have become standard\cite{mode-matching,NEGF},
and replace, to a large extent,
earlier phenomenological or parametrized procedures for quantum transport
\cite{models,Cuevas:prl:98:80}.
A widely accepted starting point is the Landauer approach which describes the electron 
transport as an elastic (and thus phase-coherent) 
scattering process of non-interacting quasi-particles.
In this approach the conductance of a nanoscopic region is determined by the 
quantum mechanical transmission probabilities of the incoming transport channels\cite{Datta:book:95}. 
The transmission matrix is calculated by mode-matching of the incoming waves in one lead
with the outgoing waves on the other lead via the intermediate scattering region\cite{mode-matching}
or, equivalently\cite{Khomyakov:prb:05}, from the one-body Green's function of the scattering region connected to the metallic leads\cite{NEGF}.
The latter is known as the non-equilibrium Green's function (NEGF) approach since 
a finite bias voltage can also be taken into account when calculating the electronic 
structure of the junction. 
One of the advantages of the NEGF approach is that it does not need to be
implemented from scratch since one can make use of standard \textit{ab initio} 
codes for the computation of the electronic structure.

It is often useful to decompose the total conductance into the contributions 
of the transport eigenchannels, first introduced by B\"uttiker\cite{Buettiker:ibmjrd:88}.
These are defined as the linear combinations of the incoming 
modes in a lead that do not mix upon reflection on the scattering region
and present a unique transmission value\cite{Brandbyge:prb:97}. 
The decomposition of the measurable total transmission
in terms of the transmissions of these
eigenchannels simplifies considerably the interpretation of the
results. Knowledge and analysis of the eigenchannel wavefunctions would, in turn,
allow one to make predictions regarding the behavior of the 
conductance upon distortions of the geometry or other perturbations of 
the scattering region\cite{Jacob:prb:05}.
Unfortunately, in the NEGF approach it is not straightforward to extract the
orbital composition of the transport eigenchannels.
Only the eigenchannel transmissions can be obtained easily in the NEGF approach 
from the non-negligible eigenvalues of the transmission matrix.
However, the associated eigenvectors turn out to be useless as obtained.
The reason is that these eigenvectors contain the contributions to the 
eigenchannel wavefunctions of the atomic orbitals at one of the borders
of the scattering region immediately connected to the leads, but not inside.



Here we present a method for analyzing the orbital contributions to the transport eigenchannels
at an arbitrary cross-section of a nanoscopic conductor
by calculating the transmission matrix projected to that cross-section.
Our approach generalizes previous work by Cuevas et al.\cite{Cuevas:prl:98:80} for tight-binding-type 
Hamiltonians to 
non-orthogonal atomic orbitals basis sets as those commonly used in quantum chemistry packages. 
An alternative approach to investigate the contributions of certain atomic or molecular orbitals
to the conductance consists in directly removing the respective orbitals from the basis set
\cite{Thygesen:prl:05:94:3}.




\section{Method}
\begin{figure}
  \includegraphics[scale=0.5]{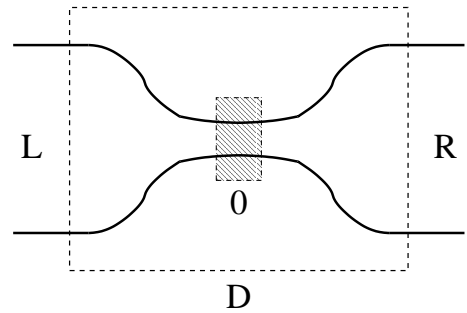}
  \caption{Sketch of the scattering problem. 
    L: Left lead. D: device. R: right Lead.
    0: cross-section of interest.}
  \label{fig:constriction}
\end{figure}

First we divide the system under study into 3 parts: The left lead (L), 
the right lead (R), and the intermediate region called device (D) from now on, 
where only elastic scattering takes place.
Figure \ref{fig:constriction}, shows a sketch of a 
nano-constriction connecting two bulk leads.
We assume that the leads are coupled only to the constriction but not to each other.
The Hamiltonian describing this situation is then given by the matrix
\begin{equation}
  \label{eq:HLDR}
  {\bf H} = \left( 
    \begin{array}{ccc}
      {\bf H}_{\rm L}  & {\bf H}_{\rm LD} & {\bf 0}      \\
      {\bf H}_{\rm DL} & {\bf H}_{\rm D}  & {\bf H}_{\rm DR} \\
      {\bf 0}          & {\bf H}_{\rm RD} & {\bf H}_{\rm R} 
    \end{array}
  \right).
\end{equation}
Since many density functional theory (DFT) codes work in non-orthogonal basis sets,
we also allow explicitly for overlap between atomic-orbitals
given by the following overlap matrix:
\begin{equation}
  \label{eq:SLDR}
  {\bf S} = \left( 
    \begin{array}{ccc}
      {\bf S}_{\rm L} & {\bf S}_{\rm LD} & {\bf 0}      \\
      {\bf S}_{\rm DL}& {\bf S}_{\rm D}  & {\bf S}_{\rm DR} \\
      {\bf 0}         & {\bf S}_{\rm RD} & {\bf S}_{\rm R} 
    \end{array}
  \right).
\end{equation}

The standard approach to calculate the conductance is to calculate the 
self-energies of the leads from the Green's functions of the isolated leads,
i.e., for the left lead ${\bf\Sigma}_{\rm L}(E) = ({\bf H}_{\rm DL}-E {\bf S}_{\rm DL}) {\bf g}_{\rm L}(E) ({\bf H}_{\rm LD}-E {\bf S}_{\rm LD})$ 
where ${\bf g}_{\rm L}(E) = (E {\bf S}_{\rm L} - {\bf H}_{\rm L})^{-1}$
is the Green's function of the isolated left lead and analogously for the right lead.
From this we can calculate the Green's function of the device:
\begin{equation}
  \label{eq:GD}
  {\bf G}_{\rm D}(E) = (E {\bf S}_{\rm D} - {\bf H}_{\rm D} - {\bf \Sigma}_{\rm L}(E) - {\bf \Sigma}_{\rm R}(E))^{-1},
\end{equation}
which, in turn, allows us to calculate the (hermitian) transmission matrix 
\begin{equation}
  \label{eq:TCarolis}
  {\bf T}(E) = {\bf \Gamma}_{\rm L}(E)^{1/2} {\bf G}_{\rm D}^\dagger(E) {\bf \Gamma}_{\rm R}(E) {\bf G}_{\rm D}(E)  {\bf \Gamma}_{\rm L}(E)^{1/2},
\end{equation}
where ${\bf\Gamma}_{\rm L}=i({\bf\Sigma}_{\rm L}-{\bf\Sigma}_{\rm L}^\dagger)$ and 
${\bf\Gamma}_{\rm R}=i({\bf\Sigma}_{\rm R}-{\bf\Sigma}_{\rm R}^\dagger)$.
Typically the leads are only connected to the left and right borders of 
the device and are sufficiently far away from the scattering region so that they
can be described by a bulk electronic structure.
From the structure of eq. (\ref{eq:TCarolis}) it follows that only 
the submatrix of ${\bf T}$ representing the subspace of the device 
immediately connected to one of the leads are non-zero. 
Thus the eigenvectors obtained by diagonalizing ${\bf T}$ only contain 
the atomic orbital contributions to the eigenchannels at one border of 
the device region but not at the center where the resistance is ultimately 
determined.


To investigate the orbital nature of the eigenchannels at an arbitrary part of the device
we can simply calculate the transmission matrix associated to this part.
By choosing this region to be a cross-section, like that indicated in Fig. \ref{fig:constriction},
current conservation guarantees that the so-calculated conductance is \emph{approximately} equal
to the conductance calculated from the transmission matrix of the whole device.
We want to emphasize here that this is really only approximately true for a Hamiltonian beyond 
the tight-binding approximation since hoppings between atoms on both sides beyond the selected 
region are neglected. Of course this approximation becomes better the thicker the chosen cross-section is.
We proceed by further subdividing the device region. The cross-section of interest will be referred to 
as 0 while the regions on either side will be denoted as l and r, respectively:
\begin{eqnarray}
  \label{eq:HD}
  {\bf H}_{\rm D} &=& \left(
    \begin{array}{ccc}
      {\bf h}_{\rm l}  & {\bf h}_{\rm l0} & {\bf h}_{\rm lr} \\
      {\bf h}_{\rm 0l} & {\bf h}_{\rm 0}  & {\bf h}_{\rm 0r} \\
      {\bf h}_{\rm rl} & {\bf h}_{\rm r0} & {\bf h}_{\rm r} 
    \end{array}
  \right)
  \\
  \nonumber \\
  \label{eq:SD}
  {\bf S}_{\rm D} &=& \left(
    \begin{array}{ccc}
      {\bf s}_{\rm l } & {\bf s}_{\rm l0} & {\bf s}_{\rm lr} \\
      {\bf s}_{\rm 0l} & {\bf s}_{\rm 0}  & {\bf s}_{\rm 0r} \\
      {\bf s}_{\rm rl} & {\bf s}_{\rm r0} & {\bf s}_{\rm r} 
    \end{array}.
  \right)
\end{eqnarray}
As mentioned above we will neglect the hoppings (and overlaps) between the 
left and right layers outside the region of interest so we set 
${\bf h}_{\rm lr} = {\bf h}_{\rm rl} = {\bf s}_{\rm lr} = {\bf s}_{\rm rl} = 0$.
With this approximation the Green's function matrix of the cross-section $0$
can be written as
\begin{equation}
  \label{eq:G0}
  {\bf G}_{\rm 0}(E) =( E {\bf s}_{\rm 0} - {\bf h}_{\rm 0} - {\bf\Sigma}^\prime_{\rm l}(E) - {\bf\Sigma}^\prime_{\rm r}(E) )^{-1}.
\end{equation}
The self-energy matrices representing the coupling to the left and right lead, 
${\bf\Sigma}^\prime_{\rm l}(E) = ({\bf h}_{\rm 0l}-E {\bf s}_{\rm 0l}) {\bf g}^\prime_{\rm l}(E) ({\bf h}_{\rm l0}-E {\bf s}_{\rm l0})$ and 
${\bf\Sigma}^\prime_{\bf r}(E) = ({\bf h}_{\rm 0r}-E {\bf s}_{\rm 0r}) {\bf g}^\prime_{\rm r}(E) ({\bf h}_{\rm r0}-E {\bf s}_{\rm r0})$, 
are given by the Green's function of the left layer $\rm l$ connected only to the left 
lead $\rm L$ and the right layer $\rm r$ connected only to the right lead $\rm R$, respectively:
\begin{eqnarray}
  {\bf g}^\prime_{\rm l}(E) &=& ( E {\bf s}_{\rm l} - {\bf h}_{\rm l} - {\bf \Sigma}_{\rm L}(E) )^{-1} \\
  {\bf g}^\prime_{\rm r}(E) &=& ( E {\bf s}_{\rm r} - {\bf h}_{\rm r} - {\bf \Sigma}_{\rm R}(E) )^{-1}. 
\end{eqnarray}
The \emph{reduced transmission matrix} (RTM) with respect to the chosen cross-section is now given by
\begin{equation}
  {\bf T}^\prime(E) = {\bf\Gamma}_{\rm l}^\prime(E)^{1/2} {\bf G}_0^\dagger(E) {\bf\Gamma}_{\rm r}^\prime(E) {\bf G}_0(E) {\bf\Gamma}_{\rm l}^\prime(E)^{1/2}
\end{equation}
with ${\bf\Gamma}_{\rm l}^\prime=i({\bf\Sigma}_{\rm l}^\prime-{{\bf\Sigma}_{\rm l}^\prime}^\dagger)$ and ${\bf\Gamma}_{\rm r}^\prime=i({\bf\Sigma}_{\rm r}^\prime-{{\bf\Sigma}_{\rm r}^\prime}^\dagger)$.
Diagonalizing ${\bf T}^\prime(E)$ now yields the contribution of the atomic orbitals within the cross-section $0$
to the eigenchannels. 

\section{Results and Discussion}
In the following we apply the above described method to analyze the orbital nature of the conducting channels
of Ni nanocontacts which have recently attracted a lot of interest because of their apparently high magnetoresistive properties
\cite{Ni-experiments}.
We consider the nanocontact to consist of two ideal pyramids facing each other along the (001) direction
and with the two tip atoms being 2.6 \r A apart. Bulk atomic distances (2.49 \r A) and perfect crystalline order are assumed for each pyramid. 
Just as in our previous work on Ni nanocontacts\cite{Jacob:prb:05} we 
perform \textit{ab initio} quantum transport calculations for this idealized geometry. To this end
we use our code ALACANT (ALicante Ab initio Computation Applied to NanoTransport). The electronic structure is computed at the 
DFT local spin density approximation 
level with a minimal basis set and the electrodes are described by means of
a semi-empirical tight-binding Bethe lattice model.

\begin{figure}
  \includegraphics[width=0.65\linewidth]{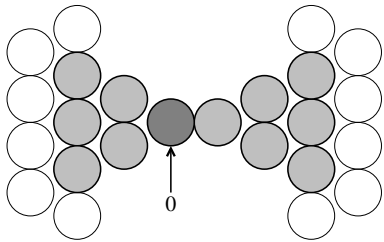}
  \caption{Sketch of Ni nanocontact consisting
    of two pyramids facing each other along the (001) 
    direction with the two tip atoms forming a dimer bridge.
    The device region (grey circles) consists of 28 Ni atoms
    and the left and right electrodes (empty circles) are modelled by Bethe lattices
    with appropriate tight-binding parameters to reproduce Ni Bulk DOS.}
  \label{fig:Ni28}
\end{figure}

\begin{figure}
  \includegraphics[width=\linewidth]{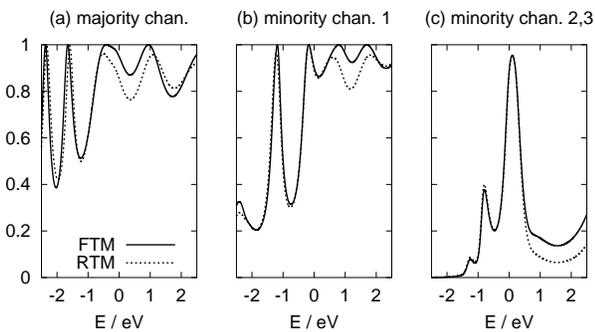}
  \caption{  
    Transmission functions of open transport channels for the Ni nanocontact 
    sketched in Fig. \ref{fig:Ni28} as calculated from the FTM
    ${\bf T}(E)$ (solid line) and from the RTM ${\bf T}^\prime(E)$ (dashed lines).
    (a) shows the only contributing M channel and (b)-(c) the three m
    channels. See text for further discussion.
  }
  \label{fig:Ni28-channels}
\end{figure}

As indicated in Fig. \ref{fig:Ni28} we calculate the RTM ${\bf T}^\prime(E)$ 
for one of the tip atoms of the contact (labelled with 0) and diagonalize it
to obtain the eigenchannels and the corresponding transmissions projected
on the tip atom.
In Fig. \ref{fig:Ni28-channels} we compare the individual channel transmissions 
calculated on the one hand from the full transmission matrix (FTM) ${\bf T}(E)$ 
and on the other hand from the RTM  ${\bf T}^\prime(E)$.
Though the electron hopping between regions l and r of the contact has been neglected
in calculating the RTM the so calculated channel transmissions approximate very well 
those calculated using the FTM so that it is very easy to relate the RTM channel 
transmissions with the FTM channel transmission.
This shows that the hopping between the regions l and r on both sides of the tip atom
is almost negligible. Only for the one majority (M) channel we see a small deviation
near the Fermi energy indicating that here 2nd neighbour hopping contributes 
to the transmission of that channel. 
As the eigenvectors of the RTM (see Table \ref{tab:1}) reveal, this channel is mainly s-type.
Since s-electrons are strongly delocalized there is a small but finite contribution 
from second-neighbour hopping explaining the deviation between the FTM and RTM transmission
in that channel.
The first minority (m) channel is also mainly s-type but now it is hybridized with 
$\rm d_{3z^2-r^2}$ and $\rm p_z$ orbitals.
The other two m channels are degenerate and mainly $\rm d_{xz}$- and
$\rm d_{yz}$-type strongly hybridized with $\rm p_x$- and $\rm p_y$-orbitals, respectively.

\begin{table}
  \begin{tabular}{l|c|c|c|c}
    AO                 & majority & minority 1 & minority 2 & minority 3 \\
    \hline 
    s                  & 97\% & 62\%  &  0   &  0   \\
    $\rm p_x$          &  0   & 0     & 28\% &  0   \\
    $\rm p_y$          &  0   & 0     &  0   & 28\% \\
    $\rm p_z$          &  3\% & 23\%  &  0   &  0   \\
    $\rm d_{3z^2-r^2}$ &  0   & 15\%  &  0   &  0   \\
    $\rm d_{xz}$       &  0   & 0     & 72\% &  0   \\
    $\rm d_{yz}$       &  0   & 0     &  0   & 72\% \\
    $\rm d_{x^2-y^2}$  &  0   & 0     &  0   &  0   \\
    $\rm d_{xy}$       &  0   & 0     &  0   &  0   \\
  \end{tabular}
  \caption{Eigenvectors of the RTM at the Fermi level
    for the contact sketched in Fig. \ref{fig:Ni28}.
    Each column gives the weights of the atomic orbitals 
    (AO) given in the left column on the tip atom in each 
    eigenchannel shown in Fig. \ref{fig:Ni28-channels}.
  }
  \label{tab:1}
\end{table}

As discussed in our previous work\cite{Jacob:prb:05} the five 
d-type transport channels for the m electrons available in the perfect Ni chain\cite{Smogunov:ss:02}
are easily blocked in a contact with a realistic geometry like that in Fig. \ref{fig:Ni28}
because the d-orbitals are very sensitive to geometry. We have refered to this
as \emph{orbital blocking}.
It is not so surprising that the $\rm d_{x^2-y^2}$- and $\rm d_{xy}$-channel 
which are very flat bands just touching the Fermi level in the perfect chain 
are easily blocked in a realistic contact geometry. These bands
represent strongly localized electrons which are easily scattered in geometries with low symmetry.
Interestingly, even the $\rm d_{3z^2-r^2}$-channel, which for the perfect chain is a very broad band 
crossing the Fermi level at half band width, does not contribute to the conduction as our eigenchannel analysis shows.
This channel is blocked because the $\rm d_{3z^2-r^2}$-orbital lying along
the symmetry axis of the contact is not ``compatible'' with the geometry 
of the two pyramids.
On the other hand the $\rm d_{xz}$- and $\rm d_{yz}$-channels are both open 
in that geometry because their shape is compatible with the pyramid geometry of the
contacts.
This illustrates how the geometry of a contact can effectively block (or open) channels 
composed of very directional orbitals. Of course, for different geometries we can expect 
different channels to be blocked or opened.

\begin{figure}
  \includegraphics[width=0.6\linewidth]{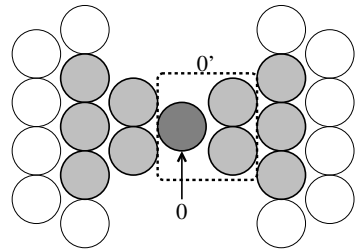}
  \caption{Sketch of Ni nanocontact (27 atoms). As in Fig. \ref{fig:Ni28}
    the contact consists of two pyramids along the (001) direction but 
    now both pyramids share the same atom at the tip.
    The device region (grey circles) consists of 27 Ni atoms
    and the left and right electrodes (empty circles) are modelled by Bethe lattices
    with appropriate tight-binding parameters to reproduce Ni bulk DOS.
  }
  \label{fig:Ni27}
\end{figure}

\begin{figure}
    \includegraphics[width=\linewidth]{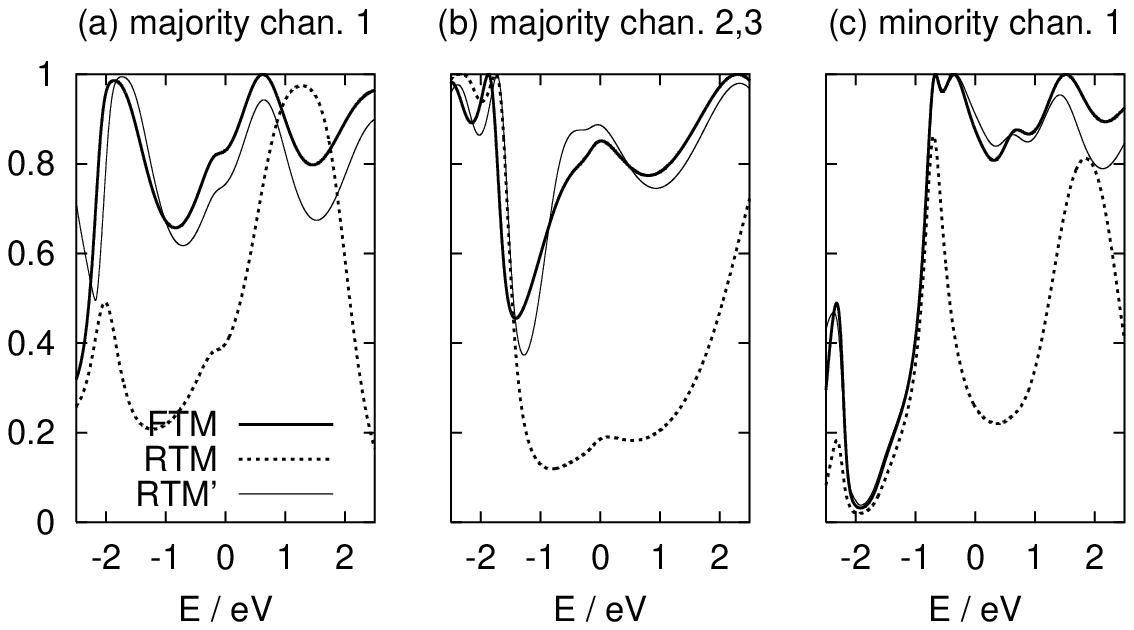} \\
    \includegraphics[width=\linewidth]{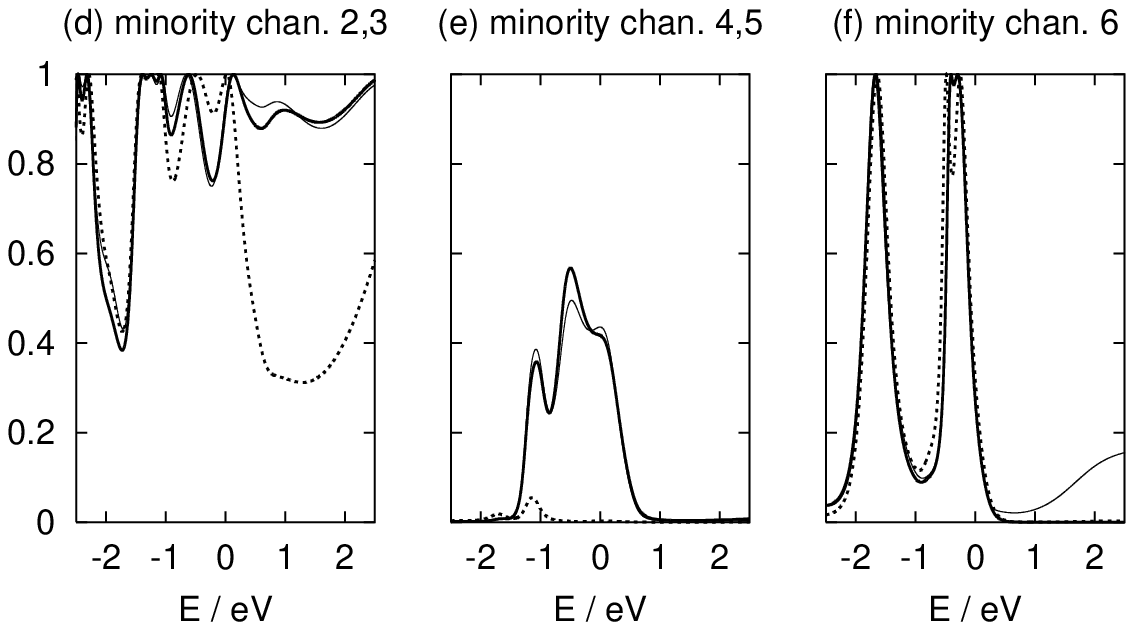} 
  \caption{  
    Transmission functions of open transport channels for the Ni nanocontact 
    sketched in Fig. \ref{fig:Ni27} as calculated from the FTM ${\bf T}(E)$ 
    (solid lines) and from the RTM ${\bf T}^\prime(E)$ (dashed lines).
    The RTM transmissions calculated for the cross-section labeled with 0$^\prime$
    in Fig. \ref{fig:Ni27} are given by the thin solid curves (labeled RTM$^\prime$).
  }
  \label{fig:Ni27-channels}
\end{figure}

Obviously, the approximation made in the calculation of the RTM
becomes worse the bigger the hopping between the regions l and r is.
For example, in the contact geometry shown in Fig. \ref{fig:Ni27}
electron hopping from the layers immediately connected to the central atom 
(labeled 0) is certainly bigger than in the geometry of Fig. \ref{fig:Ni28}.
Indeed, Fig. \ref{fig:Ni27-channels} shows that
for almost all channels the RTM transmissions  differ appreciably from FTM transmissions, making it difficult in some cases to relate them to each other.
Fortunately, we can judge by exclusion which
RTM transmission relates to which FTM transmission since for the
other channels at least the RTM transmission function mimics the overall
behaviour of the FTM transmission function.
However, for more complicated situations it might be impossible
to match the RTM transmission with the FTM transmission for all channels.
The cure to this problem is obvious: One has to choose a bigger 
cross-section, i.e., add an atomic layer to the cross-section 
so that the hopping between l and r becomes small again.
If we choose, e.g., the cross-section labeled with 0$^\prime$ in 
Fig. \ref{fig:Ni27} (including the atomic layer to the right of the central atom) 
the so calculated RTM transmissions now approximate very well the FTM transmissions as
can be seen in Fig. \ref{fig:Ni27-channels}.

\section{Conclusions}
In summary, we have shown how 
to obtain the orbital contributions to the eigenchannels at an arbitrary
cross-section of a nanoscopic conductor. 
The method has been implemented into our \textit{ab initio} quantum 
transport program ALACANT
and we have illustrated the method by exploring the orbital nature
of the eigenchannels of a Ni nanocontact. 
The method works very well when the chosen cross-section is thick 
enough so that hopping from the layers left and right to the 
cross-section becomes negligible.
Hence in some cases an additional atomic layer has to be included
to the cross-section we are actually interested in.
Taking this into account the method has no limitations and can be readily applied
to \textit{ab initio} transport calculations in all types of 
nanocontacts\cite{Fernandez:prb:05} and molecular junctions.

\section*{Acknowledgements}
We thank J. Fern\'andez-Rossier and C. Untiedt for fruitful discussions.
DJ  acknowledges financial support from MECD under grant UAC-2004-0052.
JJP acknowledges financial support from Grant No. MAT2002-04429-C03 (MCyT) and from 
University of Alicante.


\bibliography{matcon}
\bibliographystyle{apsrev}
\end{document}